\documentclass[9pt]{lapreprint}

\usepackage[misc]{ifsym}
\usepackage{orcidlink}

\makeatletter
\if@biorxiv\else\if@medrxiv\else\if@arxiv\else\if@chemrxiv\else
  \renewcommand{\ppserver}{Preprint}%
\fi\fi\fi\fi
\makeatother

\usepackage{longtable,array}
\makeatletter
\def\maxwidth{\ifdim\Gin@nat@width>\linewidth\linewidth\else\Gin@nat@width\fi}
\def\maxheight{\ifdim\Gin@nat@height>\textheight\textheight\else\Gin@nat@height\fi}
\makeatother
\setkeys{Gin}{width=\maxwidth,height=\maxheight,keepaspectratio}
\makeatletter
\newsavebox\pandoc@box
\newcommand*\pandocbounded[1]{
  \sbox\pandoc@box{#1}%
  \Gscale@div\@tempa{\textheight}{\dimexpr\ht\pandoc@box+\dp\pandoc@box\relax}%
  \Gscale@div\@tempb{\linewidth}{\wd\pandoc@box}%
  \ifdim\@tempb\p@<\@tempa\p@\let\@tempa\@tempb\fi
  \ifdim\@tempa\p@<\p@\scalebox{\@tempa}{\usebox\pandoc@box}%
  \else\usebox\pandoc@box%
  \fi%
}
\makeatother
\BeforeBeginEnvironment{longtable}{\begingroup\footnotesize}
\AfterEndEnvironment{longtable}{\endgroup}

\providecommand{\tightlist}{%
\setlength{\itemsep}{0pt}\setlength{\parskip}{0pt}}

\NewDocumentCommand\citeproctext{}{}

\makeatletter
 \let\@cite@ofmt\@firstofone
 \def\@biblabel#1{}
 \def\@cite#1#2{{#1\if@tempswa , #2\fi}}
\makeatother
\newlength{\cslhangindent}
\newlength{\csllabelwidth}
\newenvironment{CSLReferences}[2] 
 {\begin{list}{}{%
  \setlength{\itemindent}{0pt}
  \setlength{\leftmargin}{0pt}
  \setlength{\parsep}{0pt}
  \ifodd #1
   \setlength{\leftmargin}{\cslhangindent}
   \setlength{\itemindent}{-1\cslhangindent}
  \fi
  \setlength{\itemsep}{#2\baselineskip}}}
 {\end{list}}

\makeatletter
\@ifpackageloaded{subfig}{}{\usepackage{subfig}}
\@ifpackageloaded{caption}{}{\usepackage{caption}}
\AtBeginDocument{%
\renewcommand*\figurename{Figure}
\renewcommand*\tablename{Table}
}
\AtBeginDocument{%

}
\newcounter{pandoccrossref@subfigures@footnote@counter}
{\end{figure}%
\addtocounter{footnote}{-\value{pandoccrossref@subfigures@footnote@counter}}
\@for\f:=\global@pandoccrossref@subfigures@footnotes\do{\stepcounter{footnote}\footnotetext{\f}}%
\gdef\global@pandoccrossref@subfigures@footnotes{}}
\@ifpackageloaded{float}{}{\usepackage{float}}
\floatstyle{ruled}
\@ifundefined{c@chapter}{\newfloat{codelisting}{h}{lop}}{\newfloat{codelisting}{h}{lop}[chapter]}
\floatname{codelisting}{Listing}

\makeatother

\title{Hate speech toward migrants on a citizen reporting platform
concentrates in neighborhoods undergoing demographic change}

\author[1,2 \Letter]{Eduardo Graells-Garrido}
\author[3]{Daniela Opitz}
\author[4]{Francisco Rowe}
\author[4]{Carmen Cabrera}

\affil[1]{Department of Computer Science, Universidad de Chile,
Santiago, Chile}
\affil[2]{National Center for Artificial Intelligence (CENIA), Santiago,
Chile}
\affil[3]{Universidad Técnica Federico Santa María, Santiago, Chile}
\affil[4]{Geographic Data Science Lab, Department of Geography and
Planning, University of Liverpool, Liverpool, United Kingdom}

\leadauthor{Graells-Garrido}
\shorttitle{Hate speech toward migrants on a citizen reporting platform}

\metadata[]{\Letter\ Correspondence}{\href{mailto:egraells@dcc.uchile.cl}{egraells@dcc.uchile.cl}}

\metadata[]{Keywords}{hate speech, migration, citizen reporting
platforms, spatial analysis, digital bordering, urban policy}
\metadata[]{Funding}{Chilean Agency of Research and Development, through
an ANID FONDECYT Grant \#1261835; National Center for Artificial
Intelligence CENIA FB210017, Basal ANID}

\begin{document}

\maketitle

\begin{abstract}
\noindent Understanding when migration generates social integration or
exclusion is a central challenge for urban communities. Existing
research has mostly relied on surveys, administrative data, or aggregate
indicators that fail to capture expressions of exclusion at fine
spatiotemporal scales. Here, we analyze over 550,000 geolocated reports
from SOSAFE (Chile's largest citizen reporting platform) to examine the
relationship between migration and hate speech in Santiago. We fine-tune
a Spanish hate speech classifier and validate it against human labels.
Reports that mention migrants are more likely to contain hate speech
than other reports. Hate speech concentrates in areas with recent
demographic change (post-2010 arrivals) rather than in established
migrant communities. The spatial analysis shows that hate speech
hotspots coincide with neighborhoods where recent migrants comprise over
a third of the population. Coldspots appear in high-education sectors
with minimal recent migration. Reports with hate speech and reports that
mention migrants receive more engagement, although their combination is
not amplified further. These results show digital bordering on a citizen
reporting platform: exclusionary discourse concentrates, and receives
more engagement, in neighborhoods undergoing recent demographic change.
\end{abstract}

\section{Introduction}\label{introduction}

International migration has become a defining feature of contemporary
societies (Castles, 2017). Its consequences are most visible in cities,
where migrants concentrate and where new arrivals can reshape public
services, labor markets, schools, public space, and perceptions of
security. Migration can foster diversity, innovation, and social
vitality, but it can also be framed through claims about disorder,
competition, and threat (Dennison \& Geddes, 2019; Hopkins, 2010;
Putnam, 2007). Understanding when migration produces hostility requires
explaining how migrant identity becomes attached to concrete urban
problems.

Two mechanisms shape attitudes toward migrants: contact between groups
can reduce prejudice (Allport, Clark, \& Pettigrew, 1954; Pettigrew \&
Tropp, 2006), while rapid demographic change can heighten perceptions of
economic, cultural, and security threat (Hopkins, 2010). These
mechanisms are well established. Yet we know less about how they operate
in practice: how residents translate concrete urban incidents into
claims about migrant groups. Surveys estimate attitudes, interviews
provide depth, and administrative data locate demographic change. But
these sources rarely capture how anti-immigration sentiment is expressed
in real time, where it clusters within cities, and whether digital
infrastructures help it circulate. This leaves a substantive gap in our
understanding of how urban problems become ethnicized or nationalized in
everyday public communication. As a result, repeated and place-specific
patterns of stigmatization that link migrant groups to urban disorder
remain poorly documented.

Digital trace data from online platforms have emerged as a complement to
traditional sources. These data capture behavior at finer spatial and
temporal scales. In urban research, they have been used to study
mobility (Chi, Lin, Chi, \& Blumenstock, 2020), social interaction (Grow
et al., 2021), and attitudes toward migrants (Calderón, Vega, \&
Herrero, 2020; Freire-Vidal, Graells-Garrido, \& Rowe, 2021; Rowe,
Mahony, Graells-Garrido, Rango, \& Sievers, 2021). Citizen reporting
platforms are a useful source for this purpose: they generate
georeferenced, timestamped text that records how residents perceive
their immediate surroundings.

Chile offers a favorable context for studying these dynamics. The
country experienced rapid demographic change after 2010. The
foreign-born population rose from around 4\% in 2017 to 8.8\% by 2024
(Instituto Nacional de Estadísticas, 2025). This migration, mainly from
Venezuela, Colombia, Haiti, and Peru, concentrated in Santiago's urban
core. Media coverage contributed to racialized representations of
certain migrant groups as sources of insecurity and disorder (Bonhomme
\& Muirhead, 2022; Scherman, Etchegaray, Pavez, \& Grassau, 2022). The
pace and recency of this shift create variation in demographic
composition across neighborhoods within the same city, which allows
comparison of areas with recent versus established migrant communities.

We use SOSAFE, Chile's largest citizen reporting platform, which allows
residents to submit georeferenced incident reports visible to neighbors
and authorities (Tironi \& Albornoz, 2021). We apply the framework of
\emph{digital bordering}, a platform-mediated form of everyday bordering
(Yuval-Davis, Wemyss, \& Cassidy, 2019): practices that mark certain
groups as out of place through digital platforms.

We examine digital bordering using approximately 553,000 geolocated
reports submitted to SOSAFE in Santiago during 2024. We combine
report-level text, hate speech classification from a fine-tuned Spanish
language model, platform engagement metrics, and spatially disaggregated
2024 Census data. This design allows us to make three contributions.
First, we measure whether explicit migrant mentions in citizen reports
are associated with hate speech. Second, we test whether hate speech
varies with local migration trajectories, distinguishing established
from recent migration. Third, we assess whether hate speech and migrant
mentions receive higher platform engagement, which would indicate that
exclusionary framings receive attention beyond the initial act of
reporting.

Our findings show that approximately 3\% of reports mention migrants
directly, and these reports are more likely to contain hate speech.
Areas with higher proportions of recent migrants (post-2010) show
elevated hate speech rates. Hate speech and migrant mentions are each
associated with higher report engagement, which indicates that
exclusionary content receives attention beyond the initial act of
reporting. In spatial terms, hate speech clusters in areas with recent
migration. Digital bordering concentrates where demographic change is
recent.

\section{Related work}\label{related-work}

Research on migration attitudes has identified multiple determinants at
individual and contextual levels. Economic factors include labor market
competition (Burns \& Gimpel, 2000; Scheve \& Slaughter, 2001) and
fiscal concerns about public services (Hanson, Scheve, \& Slaughter,
2007). Cultural factors include perceived threats to national identity
(Esses, Dovidio, Jackson, \& Armstrong, 2002) and intergroup anxiety
(Stephan \& Stephan, 1985). The effect of contact depends on its nature:
positive contact reduces prejudice (Allport et al., 1954; Pettigrew \&
Tropp, 2006), but superficial or competitive contact can reinforce it
(Brown, Brown, Jackson, Sellers, \& Manuel, 2003; Jolly \& DiGiusto,
2014). Surveys remain the dominant measurement approach, though they
suffer from social desirability bias and limited temporal granularity
(Dennison \& Dražanová, 2018). Digital trace data offer complementary
evidence, with documented tradeoffs between coverage, granularity, and
representativeness (Armstrong, Poorthuis, Zook, Ruths, \& Soehl, 2021;
Cesare, Lee, McCormick, Spiro, \& Zagheni, 2016).

Surveys capture self-reported attitudes. Digital platforms, however,
record how these attitudes manifest as hostile communication. Hate
speech targets individuals based on group characteristics (nationality,
ethnicity, gender, religion) (Keipi, Näsi, Oksanen, \& Räsänen, 2016;
Kopytowska, 2015; Pérez-Arredondo \& Graells-Garrido, 2021). Online
platforms amplify hate speech through algorithmic curation and
engagement incentives (Bucher, 2018; Pariser, 2011). Content targeting
immigrants has been documented across languages and platforms (Basile et
al., 2019; Comandini \& Patti, 2019; Pérez-Arredondo, Ivanova, \&
Graells-Garrido, 2025; Sanguinetti, Poletto, Bosco, Patti, \& Stranisci,
2018). Computational approaches to hate speech detection have evolved
from lexicon-based methods to deep learning (Fortuna \& Nunes, 2018).
Domain adaptation remains a challenge: models that are trained on
Twitter may perform differently on citizen reporting platforms with
distinct linguistic conventions.

This hostility is shaped by platform architecture. The relationship
between hate speech and engagement creates a platform governance
challenge. Studies show that emotionally charged and divisive content
receives more shares (Vosoughi, Roy, \& Aral, 2018) and that exposure to
opposing views can increase polarization (Bail et al., 2018). These
dynamics create incentives that may conflict with community cohesion
goals (Banks, Calvo, Karol, \& Telhami, 2021).

The geography of migration attitudes varies within countries.
Established immigrant communities can reduce prejudice through sustained
contact (Jolly \& DiGiusto, 2014; Kopstein \& Wittenberg, 2009). In
Chile, migration attitudes have been studied through surveys (Dammert \&
Sandoval, 2019; Lawrence, 2015) and Twitter analysis (Freire-Vidal et
al., 2021).

Border studies supply the concept that frames these practices.
Yuval-Davis et al. (2019) describe \emph{everyday bordering}: the
boundaries of belonging are re-enacted in ordinary interactions, and
residents themselves take on the work of monitoring who is in place and
who is out of place. Citizen reporting platforms give this process a
digital and georeferenced form: each complaint can record, at street
scale, who is marked as out of place and for what reason. We use digital
bordering to name this platform-mediated version of everyday bordering,
and we examine where it concentrates within a city and how much
attention it receives.

Citizen reporting platforms occupy an intermediate position between
social media and formal systems. Research on similar platforms in the
United States has documented patterns of racial profiling and
exclusionary discourse (Bloch, 2022; Choksi et al., 2024; Lambright,
2019). Research documents how platform affordances shape reporting
behavior: ``suspicious activity'' categories enable discriminatory
vigilance (Larsson, 2017; Parker \& Dodge, 2024), and reporting patterns
show class-based disparities (Iqbal, Ghafouri, Tyson, Suarez-Tangil, \&
Castro, 2023). To our knowledge, migration discourse on SOSAFE has not
been studied.

\section{Data}\label{data}

This study combines three data sources: user-generated reports from the
SOSAFE platform, hate speech classifications from a fine-tuned language
model, and demographic variables from the 2024 Chilean Census.

\begin{figure}
\centering
\pandocbounded{\includegraphics[keepaspectratio,alt={Daily number of SOSAFE reports in Santiago during 2024 (panel a), fraction of migrant mentions (panel b), and fraction of reports containing hate speech (panel c).}]{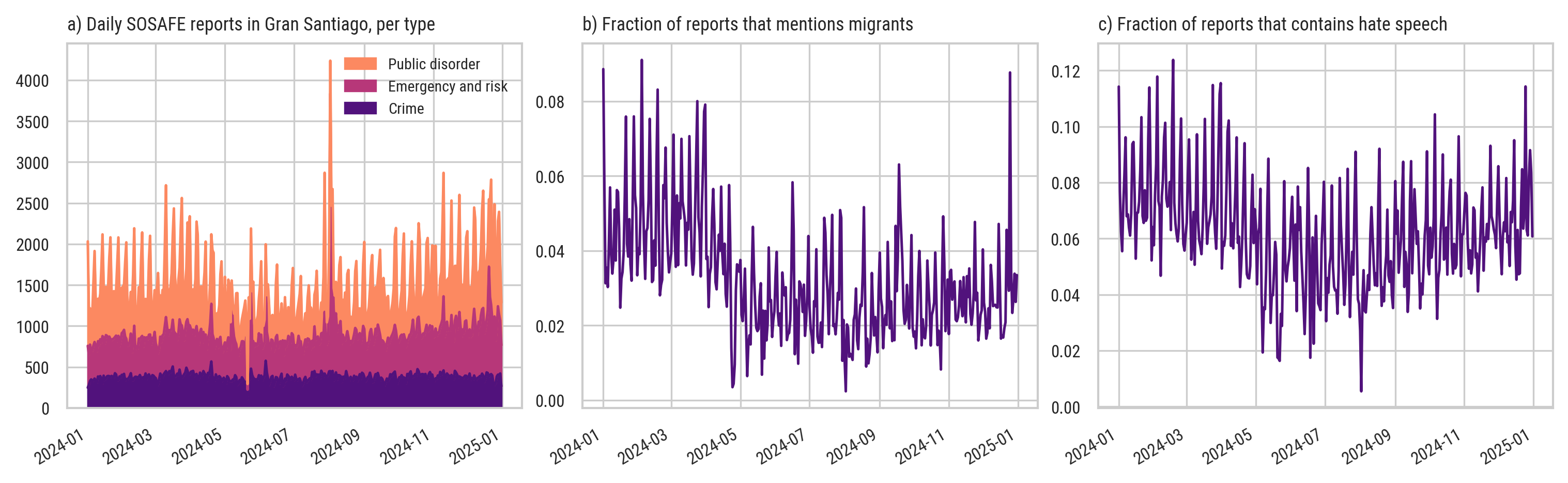}}
\caption{Daily number of SOSAFE reports in Santiago during 2024 (panel
a), fraction of migrant mentions (panel b), and fraction of reports
containing hate speech (panel c).}\label{fig:daily_reports}
\end{figure}

We collected SOSAFE reports from the platform's public map. Our dataset
consists of 553,400 reports from January 1 to December 31, 2024. Each
report contains a timestamp, geographic coordinates, user-generated
text, a numerical category selected by the user, and engagement metrics
(likes and comments). Reports are visible to nearby users, who can react
to them; we observe these reactions but not views or the ordering of the
feed, so engagement measures reception rather than algorithmic
amplification. All reports are in Spanish and located within Santiago,
Chile's capital. We excluded reports from other cities and categories
unrelated to neighborhood incidents (e.g., lost pets, service
offerings). Figure \ref{fig:daily_reports} shows the daily volume of
reports during 2024, together with the fraction of migrant mentions and
the fraction classified as hate speech. Overall reporting varies
seasonally. The fraction of migrant mentions is higher during the first
months of the year and stabilizes around 2--3\% after April, and the
hate speech fraction declines during the austral winter and rises again
toward the end of the year; the month fixed effects analysis accounts
for this seasonality (Appendix E). The spike in August corresponds to a
prolonged power outage across much of the country that generated
widespread incident reporting.

\begin{figure}
\centering
\pandocbounded{\includegraphics[keepaspectratio,alt={Distribution of SOSAFE reports by category. Left panel shows the fraction of reports in each category that mention migrants; right panel shows total report counts (log scale)}]{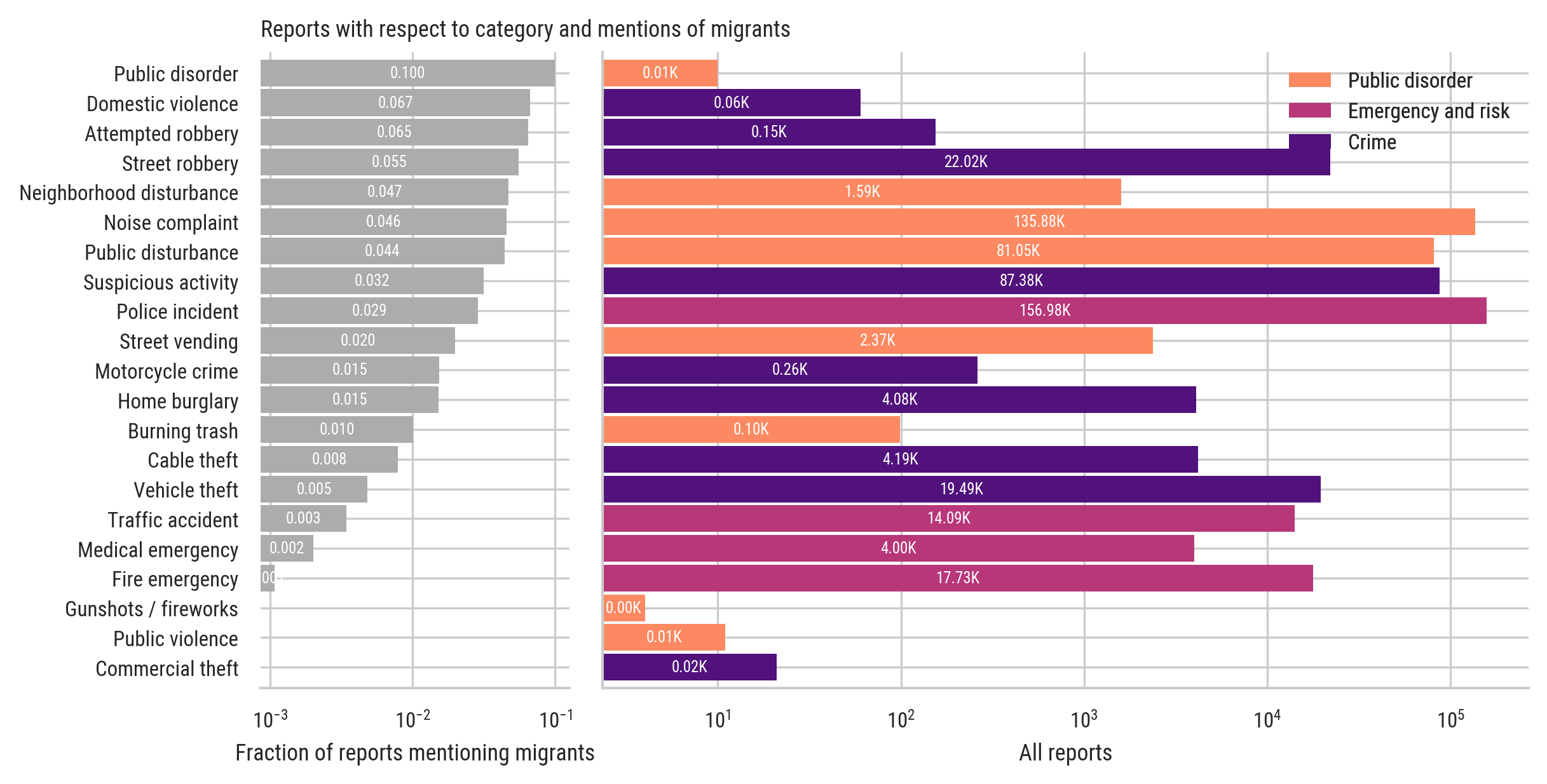}}
\caption{Distribution of SOSAFE reports by category. Left panel shows
the fraction of reports in each category that mention migrants; right
panel shows total report counts (log scale)}\label{fig:reports_category}
\end{figure}

The numerical category selected by the user has no documented meaning.
We therefore inspected the most frequent words and a sample of reports
under each value and assigned it a human-readable name that describes
the incident type (for example, street robbery, noise complaint, or
medical emergency). We gave the same name to values that describe the
same incident type, which yields 21 named categories. To ease
interpretation, we also grouped the categories into three broader types:
Crime (e.g., robbery, vehicle theft, home burglary), Emergency and Risk
(e.g., fire, medical emergency, traffic accident), and Public Disorder
(e.g., noise complaints, suspicious activity, street vending). Figure
\ref{fig:reports_category} displays the distribution of reports across
these categories and the share of migrant mentions within each category.

We identified migrant mentions through keyword matching of demonyms
(e.g., ``venezolano'', ``colombiano'', ``haitiano'') and
migration-related terminology (e.g., ``extranjero'', ``inmigrante'').
The pattern also matches common obfuscations (leetspeak such as
``v3n3z0l4n0s'') and derogatory portmanteaus. These mentions appear in
3.38\% of reports and concentrate in everyday neighborhood conflicts
rather than acute emergencies (Figure \ref{fig:reports_category}): 10\%
of ``public disorder'' reports and 6.7\% of ``domestic violence''
reports mention migrants, while emergency categories such as ``medical
emergency'' (0.2\%) and ``fire emergency'' (near 0\%) rarely do.

\begin{figure}
\centering
\pandocbounded{\includegraphics[keepaspectratio,alt={Distribution of reports by day of week (panel a) and hour of day (panel b).}]{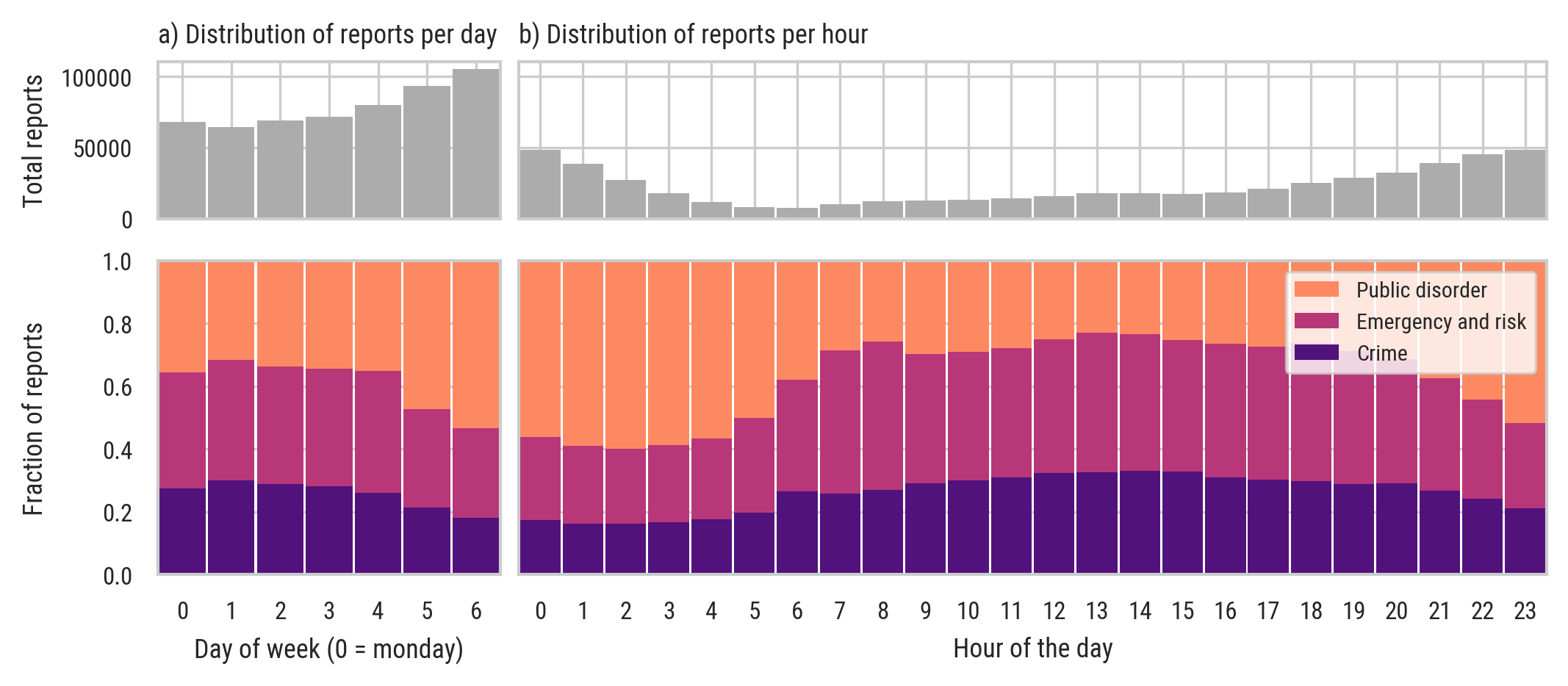}}
\caption{Distribution of reports by day of week (panel a) and hour of
day (panel b).}\label{fig:temporal}
\end{figure}

Reports are more frequent during nighttime hours (20:00--04:00) and on
weekends (Figure \ref{fig:temporal}). The composition of report types
varies across time: public disorder reports dominate during nighttime
and weekends, while crime reports constitute a larger share during
daytime hours.

To detect hate speech in reports, we fine-tuned a transformer-based
language model for Spanish. We compared two base models: RoBERTuito
(Pérez, Furman, Alonso Alemany, \& Luque, 2022), a RoBERTa model
pre-trained on Spanish tweets, and SinOdio-BETO (Dromundo, 2025a), a
BETO model (Cañete et al., 2020) trained on hate speech detection for
Latin American Spanish using the SinOdio-LATAM dataset (Dromundo,
2025b). We fine-tuned both on a combination of Chilean Twitter hate
speech data (Arango Monnar, Perez Rojas, \& Poblete Labra, 2024) and
manually labeled SOSAFE reports to adapt them to our domain (see
Appendix A for model comparison). For manual labeling, we followed the
United Nations definition of hate speech: communication that attacks or
uses pejorative or discriminatory language with reference to a person or
group based on characteristics such as religion, ethnicity, nationality,
race, or gender (United Nations, n.d.).

To prevent the classifier from learning associations between specific
nationalities and hate speech, we masked demonyms during training and
inference (see Appendix A). This forces the model to learn hostility
from surrounding context rather than from specific nationality terms. We
selected RoBERTuito fine-tuned with nationality masking because it
achieved the highest mean F1 and ROC AUC of the four model
configurations we compared (Appendix A). F1 is the harmonic mean of
precision and recall, and ROC AUC is the probability that the model
scores a random hate speech report above a random non-hate report. Under
5-fold cross-validation it reaches F1 = 0.62 \(\pm\) 0.06 and ROC AUC =
0.81 \(\pm\) 0.04; the deployed instance reaches F1 = 0.500 (95\%
bootstrap CI: {[}0.311, 0.667{]}) on its held-out test set of 100
reports (see Appendix A for model details). Appendix A also validates
the deployed classifier on a blind sample at the true prevalence of hate
speech. With this classifier, 6.5\% of reports are labeled as containing
hate speech.

We complement the SOSAFE data with the 2024 Chilean Census (Instituto
Nacional de Estadísticas, 2025). The census publishes household-level
microdata whose geographic detail stops at the municipality, and
separately publishes aggregate counts per city block. We developed a
simulated annealing approach (Kirkpatrick, Gelatt Jr, \& Vecchi, 1983)
that combines both sources: it assigns each household to an H3 hexagonal
cell (Uber Technologies, 2026) within its municipality so that the
assigned totals approximate the block-level aggregates (Appendix B). The
assignment runs at H3 resolution 8 (approximately 0.74 km\(^2\)). We use
this resolution because its cells approximate the scale of a
neighborhood; we test sensitivity to a coarser resolution in Appendix E.

In the census, migrants specified a period of arrival to the country. We
created two migration variables per spatial unit from those answers:

\begin{itemize}
\tightlist
\item
  \textbf{Established migration}: Fraction of population that arrived
  before 2010.
\item
  \textbf{Recent migration}: Fraction of population that arrived since
  2010.
\end{itemize}

\begin{figure}
\centering
\pandocbounded{\includegraphics[keepaspectratio,alt={Migrant population in Santiago by period of arrival (panel a) and origin distribution (panel b).}]{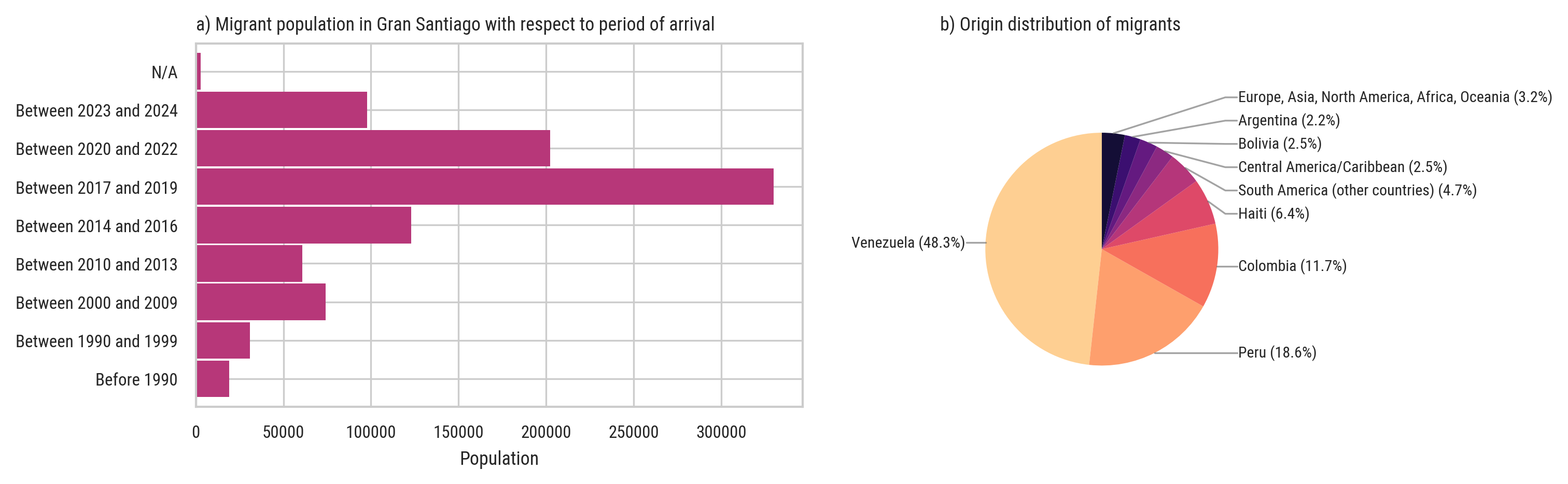}}
\caption{Migrant population in Santiago by period of arrival (panel a)
and origin distribution (panel b).}\label{fig:migration_origin}
\end{figure}

We make this distinction because Chile's migration patterns changed
substantially after 2010. Before 2010, immigration was small and came
mainly from neighboring countries. The post-2010 period saw rapid
growth: Venezuelan (Freier \& Parent, 2019), Colombian, and Haitian
migration drove this change. Figure \ref{fig:migration_origin} shows
that the largest wave arrived between 2017 and 2019. The 2020-2022
cohort was the second largest. Venezuela (48.3\%), Peru (18.6\%),
Colombia (11.7\%), and Haiti (6.4\%) account for 85\% of Santiago's
migrant population.

The census does not contain income information. However, educational
level is highly correlated with income (Crespo \& Hernandez, 2020).
Hence, to account for income level, we computed two educational
attainment variables:

\begin{itemize}
\tightlist
\item
  \textbf{High educational level}: Fraction of population with tertiary
  education (technical or university/postgraduate).
\item
  \textbf{Base educational level}: Fraction of population with primary
  or secondary education.
\end{itemize}

The two fractions are near-complementary, so we use only the
tertiary-education fraction as the income proxy in the regression models
and report both fractions in the descriptive tables (see Appendix E).

\begin{figure}
\centering
\pandocbounded{\includegraphics[keepaspectratio,alt={Spatial distribution of key variables across Santiago (H3 resolution 8). Panels show: (a) population, (b) number of reports, (c) fraction of reports with hate speech, (d-f) fraction of Crime, Emergency and Risk, and Public Disorder reports, (g) fraction with tertiary education, (h) established migration (pre-2010), and (i) recent migration (post-2010).}]{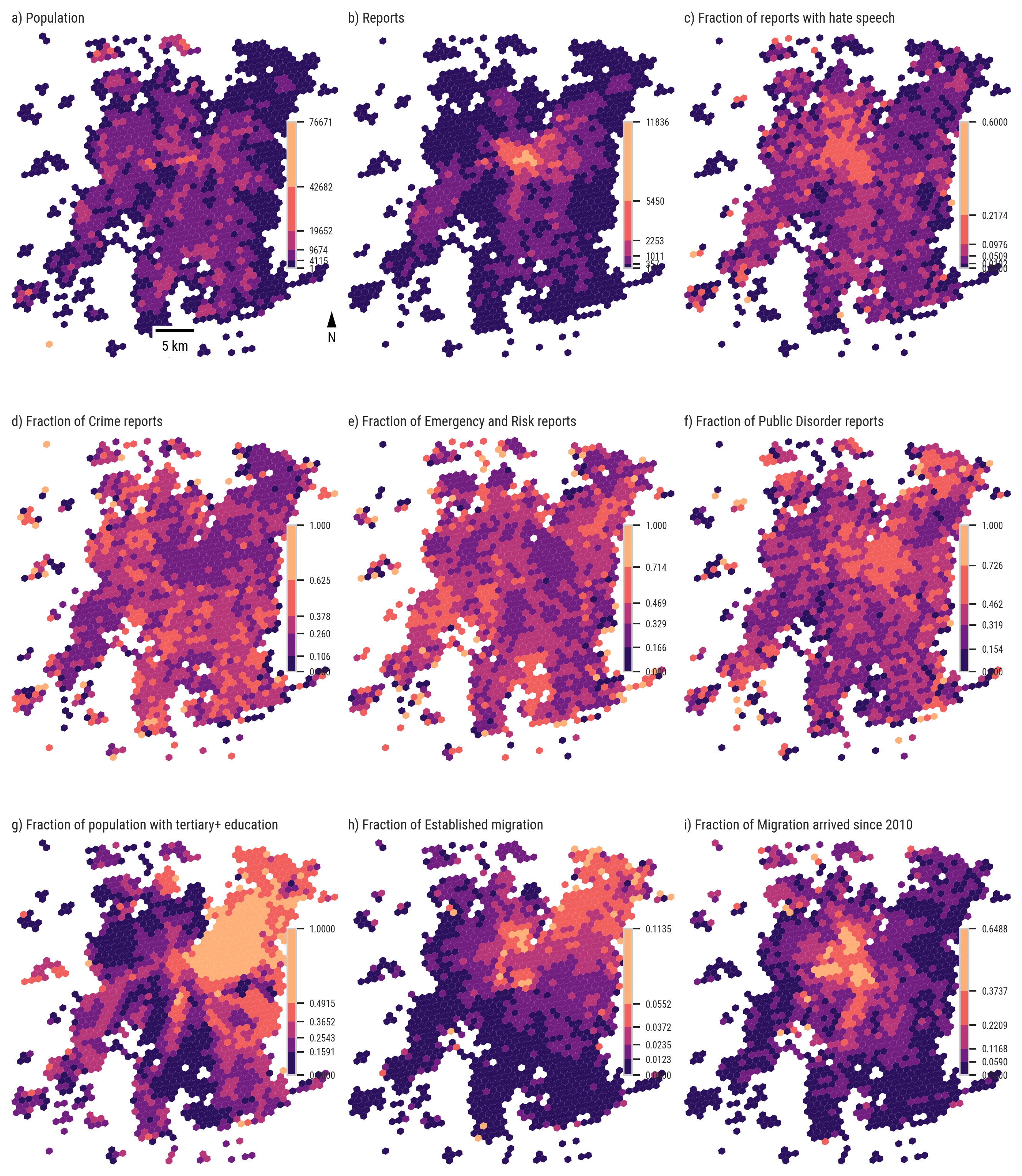}}
\caption{Spatial distribution of key variables across Santiago (H3
resolution 8). Panels show: (a) population, (b) number of reports, (c)
fraction of reports with hate speech, (d-f) fraction of Crime, Emergency
and Risk, and Public Disorder reports, (g) fraction with tertiary
education, (h) established migration (pre-2010), and (i) recent
migration (post-2010).}\label{fig:spatial}
\end{figure}

We assigned each report to an H3 hexagonal cell based on its
coordinates. We then joined census variables to reports through spatial
indexing. We excluded reports in cells with fewer than 100 total reports
to ensure stable estimates; this filter keeps 537,453 reports (97.1\% of
the dataset) for the regression analyses. Figure \ref{fig:spatial} shows
the spatial distributions for all variables. The eastern sector shows
higher education levels and lower hate speech rates. Recent migration
concentrates in the central core, where hate speech rates are also
elevated; some peripheral areas show high hate speech rates as well.

\section{Methods}\label{methods}

Our analysis has two stages. We first estimate regression models to
examine associations between migrant mentions, area characteristics,
hate speech, and engagement. Then, we apply spatial cluster analysis to
identify geographic concentrations of hate speech and examine their
demographic composition.

The first stage comprises three regression models, all estimated by
maximum likelihood via \texttt{statsmodels} (Seabold, Perktold, et al.,
2010). We model the probability that a report \textbf{contains hate
speech} using logistic regression:
\[\text{logit}(P(\text{has\_hate}_i = 1)) = \beta_0 + \beta_1 \text{migrant\_mention}_i + \sum_{k} \beta_k X_{ki} + \sum_{m} \gamma_m Z_{mj}.\]
Here \(\text{has\_hate}_i\) is a binary indicator equal to 1 when report
\(i\) is classified as hate speech, and \(\text{migrant\_mention}_i\) is
a binary indicator equal to 1 when report \(i\) mentions migrants. The
subscript \(i\) indexes reports and \(j\) indexes the H3 cell that
contains report \(i\). The terms \(X_{ki}\) are report-level covariates
(category dummies), and \(Z_{mj}\) are cell-level covariates (log
population, migration fractions, and education). We cluster standard
errors at the cell level to account for within-cell correlation (Abadie,
Athey, Imbens, \& Wooldridge, 2017). We assign cell-level covariates to
individual reports through spatial joining, so area-level associations
should be read as ecological correlations rather than individual-level
effects. The area-level covariates show low collinearity (VIF
\textless{} 2; see Appendix E).

We model \textbf{likes} and \textbf{comments} using negative binomial
regression, which accounts for overdispersion in count data (Cameron \&
Trivedi, 1986). For the expected count of likes:

\begin{align*}
\log(E[\text{likes}_i]) = \alpha_0 &+ \alpha_1 \text{has\_hate}_i + \alpha_2 \text{migrant\_mention}_i \\
&+ \alpha_3 (\text{has\_hate}_i \times \text{migrant\_mention}_i) \\
&+ \sum_{k} \beta_k X_{ki} + \sum_{m} \gamma_m Z_{mj}.
\end{align*}

The interaction term \(\alpha_3\) tests whether hate speech directed at
migrants generates different engagement compared to what would be
predicted from the main effects alone. A positive \(\alpha_3\) indicates
amplification; a negative value indicates that the combined effect is
less than additive. We estimate an analogous model for comments. We
again cluster standard errors at the cell level.

To find geographic concentrations of hate speech, we apply Local
Indicators of Spatial Association (LISA) using \texttt{PySAL} (Anselin,
1995). LISA statistics decompose global spatial autocorrelation
(measured by Moran's I) into local contributions and enable
identification of statistically significant clusters.

For each cell \(i\), the local Moran's I statistic is:
\[I_i = \frac{(x_i - \bar{x})}{\sum_j (x_j - \bar{x})^2 / n} \sum_j w_{ij}(x_j - \bar{x}),\]
where \(x_i\) is the hate speech rate in cell \(i\), \(\bar{x}\) is the
global mean, and \(w_{ij}\) are spatial weights. We create a Queen
contiguity weights matrix from the hexagonal grid, where \(w_{ij} = 1\)
if cells \(i\) and \(j\) share an edge or vertex, and \(w_{ij} = 0\)
otherwise. Weights are row-standardized so that \(\sum_j w_{ij} = 1\)
for each \(i\).

We run the analysis on the 888 cells that meet the minimum of 100
reports described above, which yields one local test per cell. We assess
statistical significance using conditional permutation inference with
999 random permutations per test. We classify cells at \(p < 0.05\) into
four cluster types based on the sign of \((x_i - \bar{x})\) and the
spatial lag \(\sum_j w_{ij}(x_j - \bar{x})\): HH (hotspots: high values
surrounded by high values), LL (coldspots), and spatial outliers (HL,
LH). After identifying clusters, we examine the distribution of
migration variables (established vs.~recent) and educational composition
across cluster types.

We assess the robustness of these analyses through re-estimation at a
coarser spatial resolution, spatial error and lag models at the cell
level, month fixed effects, LISA cluster fixed effects, a false
discovery rate correction for the local tests, Empirical Bayes smoothing
of the cell rates, re-estimation on reports without migrant mentions,
and variation of the classification threshold (Appendices C, D, and E).

\section{Results}\label{results}

We first note the unconditional association in classifier labels: 40.3\%
of reports that mention migrants contain hate speech, versus 5.4\% of
reports that do not (a relative risk of 7.48). The observed
co-occurrence of migrant mentions and hate speech is six times what
would be expected under independence (\(\chi^2 =\) 35,302, \(p <\)
0.001; this test assumes independent reports, and the regression models
below account for within-cell correlation). The rate among
migrant-mention reports is consistent with the human-labeled data, where
48\% of keyword-sampled reports that mention migrants were annotated as
hate speech. In a blind validation sample drawn at deployment
prevalence, human labels yield a relative risk of 9.7, so this
association does not come from classifier error (Appendix A).

Table \ref{tbl:hate_model} presents logistic regression results for hate
speech. Migrant mention is associated with hate speech (OR = 9.43,
\(\beta =\) 2.24, \(p <\) 0.001), controlling for other factors. At the
mean of all covariates, a migrant mention increases the predicted hate
speech probability by 9.7 percentage points (marginal effect = 0.097,
\(p <\) 0.001).

Recent migration is positively associated with hate speech (\(\beta =\)
1.54, \(p <\) 0.001): areas where recent arrivals constitute a larger
fraction of the population show elevated hate speech rates. Established
migration also shows a positive coefficient (\(\beta =\) 6.72, \(p =\)
0.001), but this variable has low variance across the city (most areas
have very few pre-2010 migrants). The two coefficients become comparable
per standard deviation of each variable: one standard deviation of
recent migration raises the log-odds of hate speech by 0.21, against
0.09 for established migration (Appendix E). The established migration
association is also fragile. It loses significance at a coarser spatial
resolution, in the cell-level spatial models, and in the model that adds
a \emph{convenio} indicator, which marks municipalities that signed a
formal agreement to integrate SOSAFE into their security operations.
Recent migration remains significant in all specifications (Appendices D
and E). The spatial cluster analysis below shows how these associations
distribute geographically.

\begin{longtable}[]{@{}lrrrr@{}}
\caption{\label{tbl:hate_model}Logistic regression results for hate
speech (\(N =\) 537,453). Pseudo \(R^2\) = 0.137. Standard errors are
clustered by hexagonal cell to account for correlation among reports in
the same cell (see Methods).}\tabularnewline
\toprule\noalign{}
Variable & Coefficient & Std. Error & \(z\) & \(p\)-value \\
\midrule\noalign{}
\endfirsthead
\toprule\noalign{}
Variable & Coefficient & Std. Error & \(z\) & \(p\)-value \\
\midrule\noalign{}
\endhead
\bottomrule\noalign{}
\endlastfoot
Intercept & -5.35 & 0.26 & -20.46 & \textless0.001 \\
Migrant mention & 2.24 & 0.04 & 57.22 & \textless0.001 \\
Emergency and risk (ref: Crime) & 0.76 & 0.05 & 15.86 &
\textless0.001 \\
Public disorder (ref: Crime) & 1.80 & 0.07 & 26.74 & \textless0.001 \\
log(Population) & 0.13 & 0.03 & 4.30 & \textless0.001 \\
Established migration & 6.72 & 2.11 & 3.18 & 0.001 \\
Recent migration (post-2010) & 1.54 & 0.21 & 7.17 & \textless0.001 \\
High educational level & -0.97 & 0.12 & -7.76 & \textless0.001 \\
\end{longtable}

Public Disorder reports show higher hate speech rates than Crime reports
(\(\beta =\) 1.80, \(p <\) 0.001). These complaints describe people and
their behavior (noise, ``suspicious'' activity, conflicts over public
space) rather than discrete criminal events, which leaves more room for
identity-based language.

Areas with more tertiary education show less hate speech (\(\beta =\)
-0.97, \(p <\) 0.001). This is consistent with the income interpretation
of the education proxy: hate speech rates are lower in higher-income
areas. We explore this pattern further in the spatial analysis.

The model's Pseudo \(R^2\) of 0.137 indicates that our covariates
capture a modest share of variance in hate speech occurrence. This is
expected: individual report content is driven by idiosyncratic factors
that area-level demographics cannot capture. We design the models to
estimate associations rather than to predict individual reports, and the
large sample provides precise coefficient estimates despite low overall
explanatory power.

Spatial associations can depend on the scale of the aggregation units, a
concern known as the modifiable areal unit problem (MAUP) (Openshaw,
1984). Our central result is a report-level association: migrant mention
and hate speech are both measured on individual reports, so their
association is defined independently of the spatial units, and
resolution enters only through the cell-level covariates (migration and
education). To confirm this empirically, we re-estimate the model at H3
resolution 7 (approximately 5.2 km\(^2\) per cell, about seven times
larger than resolution 8). The report-level coefficient for migrant
mention remains stable (\(\beta =\) 2.26 at resolution 7 vs.~\(\beta =\)
2.24 at resolution 8, \(p <\) 0.001). Recent migration retains its
positive association (\(\beta =\) 1.98, \(p <\) 0.001), though with
larger standard errors due to the coarser aggregation. Tertiary
education also remains significant (\(\beta =\) -0.88, \(p <\) 0.001),
while established migration loses significance, as expected given
greater within-cell heterogeneity (Appendix E).

We also test for residual spatial autocorrelation. The Moran's I
statistic for cell-level Pearson residuals is 0.233 (\(p <\) 0.001).
This indicates that the logistic model does not fully capture spatial
structure. As a robustness check, we estimate a spatial error model at
the cell level using \texttt{spreg} (Anselin \& Rey, 2014). This model
runs on aggregated cells, so its dependent variable is the cell hate
speech rate and its migrant mention regressor is the fraction of reports
in the cell that mention migrants. For this reason its coefficients are
not comparable in magnitude to the report-level model above. The spatial
error parameter (\(\lambda =\) 0.36, \(p <\) 0.001) confirms significant
spatial dependence. In this model the migrant mention rate (\(\beta =\)
0.56, \(p <\) 0.001) and recent migration (\(\beta =\) 0.05, \(p <\)
0.001) remain significant, which means our main findings are not
artifacts of unmodeled spatial dependence (Appendix E).

Month fixed effects do not alter the key coefficients (migrant mention
\(\beta =\) 2.23 vs.~2.24 in the base model), though hate speech rates
are lower during the austral winter months (May--September) relative to
the summer baseline (Appendix E).

We now examine whether exclusionary content achieves wider circulation
within the platform.

\begin{figure}
\centering
\pandocbounded{\includegraphics[keepaspectratio,alt={Relationship between hate speech and engagement. Panels (a) and (c) show the fraction of reports containing hate speech across engagement levels, comparing reports that mention migrants (orange) versus those that do not (purple). Panels (b) and (d) show the distribution of reports across engagement levels.}]{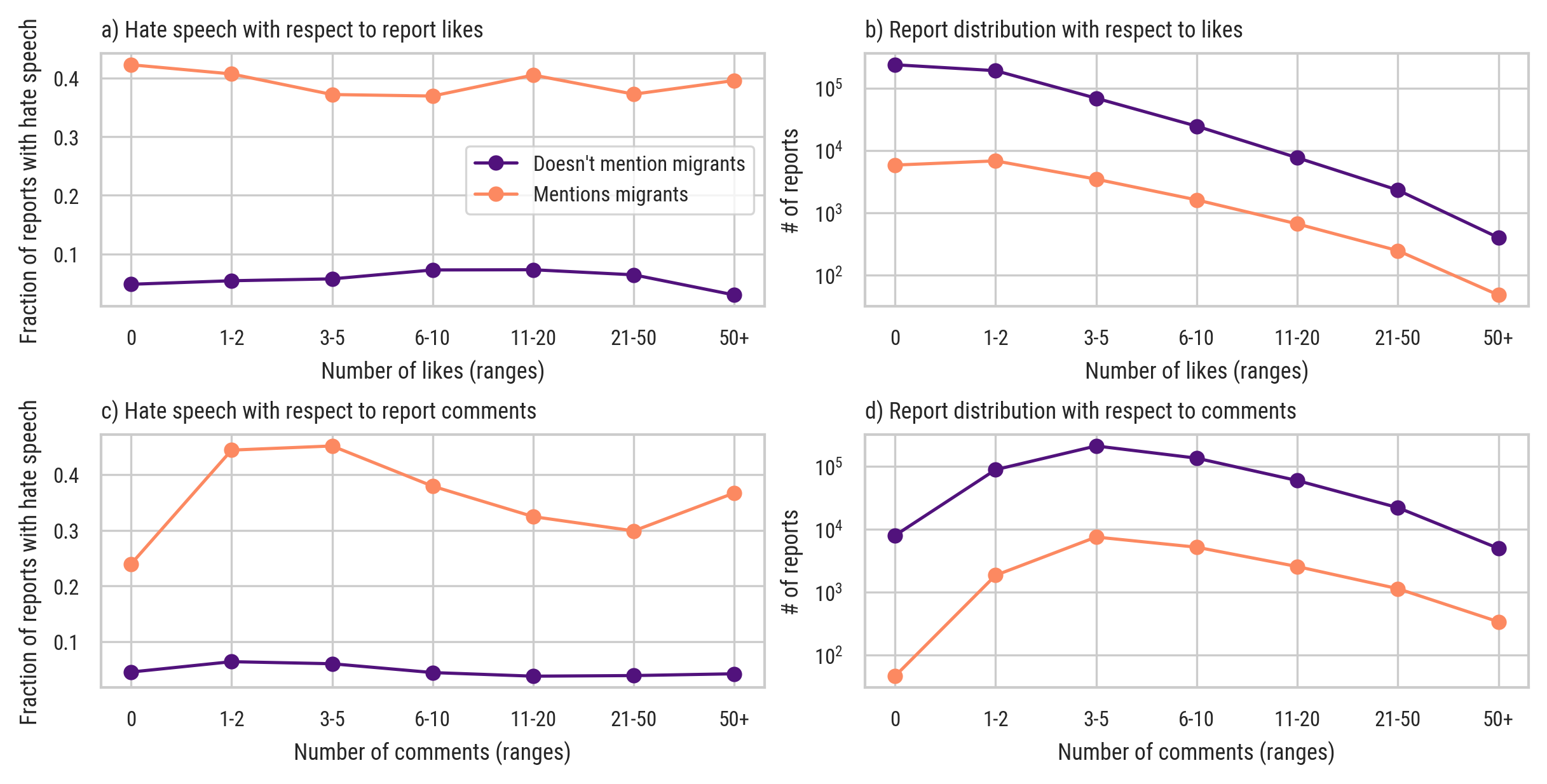}}
\caption{Relationship between hate speech and engagement. Panels (a) and
(c) show the fraction of reports containing hate speech across
engagement levels, comparing reports that mention migrants (orange)
versus those that do not (purple). Panels (b) and (d) show the
distribution of reports across engagement levels.}\label{fig:engagement}
\end{figure}

Reports that mention migrants show higher hate speech rates across all
engagement levels (Figure \ref{fig:engagement}). Most reports receive
zero or 1--2 likes, while comments concentrate at 3--5 per report; both
distributions decline at higher engagement levels (panels b, d). Reports
that mention migrants are between one and two orders of magnitude less
numerous than other reports, depending on the engagement bin. But hate
speech rates show the opposite pattern: reports with migrant mentions
show rates of 37--42\% across like levels and 24--45\% across comment
levels, compared to 3--7\% and 4--6\% for other reports.

Table \ref{tbl:engagement_model} presents the engagement model results.
For likes, hate speech is associated with more engagement (\(\beta =\)
0.37, \(p <\) 0.001), and so is migrant mention (\(\beta =\) 0.58,
\(p <\) 0.001). However, the interaction term is negative and
significant (\(\beta =\) -0.23, \(p <\) 0.001). Reports that combine
hate speech and a migrant mention receive more likes than reports with
either feature alone, but less than the sum of the two main effects
would predict. The result suggests either a ceiling effect or segmented
audiences who respond to one signal but not their combination.

\begin{longtable}[]{@{}
  >{\raggedright\arraybackslash}p{(\linewidth - 12\tabcolsep) * \real{0.1333}}
  >{\raggedleft\arraybackslash}p{(\linewidth - 12\tabcolsep) * \real{0.1200}}
  >{\raggedleft\arraybackslash}p{(\linewidth - 12\tabcolsep) * \real{0.1333}}
  >{\raggedleft\arraybackslash}p{(\linewidth - 12\tabcolsep) * \real{0.1200}}
  >{\raggedleft\arraybackslash}p{(\linewidth - 12\tabcolsep) * \real{0.1600}}
  >{\raggedleft\arraybackslash}p{(\linewidth - 12\tabcolsep) * \real{0.1733}}
  >{\raggedleft\arraybackslash}p{(\linewidth - 12\tabcolsep) * \real{0.1600}}@{}}
\caption{\label{tbl:engagement_model}Negative binomial regression
results for engagement (\(N =\) 537,453). Standard errors clustered by
hexagonal cell (see Methods). Likes model Pseudo \(R^2\) = 0.024;
Comments model Pseudo \(R^2\) = 0.032.}\tabularnewline
\toprule\noalign{}
\begin{minipage}[b]{\linewidth}\raggedright
Variable
\end{minipage} & \begin{minipage}[b]{\linewidth}\raggedleft
Likes \(\beta\)
\end{minipage} & \begin{minipage}[b]{\linewidth}\raggedleft
Likes SE
\end{minipage} & \begin{minipage}[b]{\linewidth}\raggedleft
Likes \(p\)
\end{minipage} & \begin{minipage}[b]{\linewidth}\raggedleft
Comments \(\beta\)
\end{minipage} & \begin{minipage}[b]{\linewidth}\raggedleft
Comments SE
\end{minipage} & \begin{minipage}[b]{\linewidth}\raggedleft
Comments \(p\)
\end{minipage} \\
\midrule\noalign{}
\endfirsthead
\toprule\noalign{}
\begin{minipage}[b]{\linewidth}\raggedright
Variable
\end{minipage} & \begin{minipage}[b]{\linewidth}\raggedleft
Likes \(\beta\)
\end{minipage} & \begin{minipage}[b]{\linewidth}\raggedleft
Likes SE
\end{minipage} & \begin{minipage}[b]{\linewidth}\raggedleft
Likes \(p\)
\end{minipage} & \begin{minipage}[b]{\linewidth}\raggedleft
Comments \(\beta\)
\end{minipage} & \begin{minipage}[b]{\linewidth}\raggedleft
Comments SE
\end{minipage} & \begin{minipage}[b]{\linewidth}\raggedleft
Comments \(p\)
\end{minipage} \\
\midrule\noalign{}
\endhead
\bottomrule\noalign{}
\endlastfoot
Intercept & -1.28 & 0.36 & \textless0.001 & 2.24 & 0.15 &
\textless0.001 \\
Migrant mention & 0.58 & 0.03 & \textless0.001 & 0.35 & 0.02 &
\textless0.001 \\
Emergency and risk & -0.15 & 0.01 & \textless0.001 & 0.10 & 0.01 &
\textless0.001 \\
Public disorder & -0.94 & 0.02 & \textless0.001 & -0.53 & 0.02 &
\textless0.001 \\
Has hate speech & 0.37 & 0.02 & \textless0.001 & 0.06 & 0.01 &
\textless0.001 \\
Hate \(\times\) Migrant mention & -0.23 & 0.04 & \textless0.001 & -0.10
& 0.03 & 0.003 \\
log(Population) & 0.22 & 0.04 & \textless0.001 & 0.05 & 0.02 & 0.005 \\
Established migration & -9.44 & 1.75 & \textless0.001 & -5.48 & 1.03 &
\textless0.001 \\
Recent migration & 0.11 & 0.21 & 0.602 & -0.10 & 0.10 & 0.322 \\
High educational level & 1.10 & 0.11 & \textless0.001 & -1.14 & 0.07 &
\textless0.001 \\
\end{longtable}

For comments, patterns differ. Hate speech shows a modest positive
effect (\(\beta =\) 0.06, \(p <\) 0.001), while migrant mentions
increase comments (\(\beta =\) 0.35, \(p <\) 0.001). The interaction is
negative and significant (\(\beta =\) -0.10, \(p =\) 0.003). The pattern
is again sub-additive, and here the combined effect falls below the
migrant mention effect alone: hateful content about migrants does not
receive additional commentary beyond what migrant mentions produce.

Established migration is negatively associated with both likes
(\(\beta =\) -9.44) and comments (\(\beta =\) -5.48), while recent
migration shows no significant effect on either engagement metric.
Higher tertiary education is associated with more likes (\(\beta =\)
1.10) but fewer comments (\(\beta =\) -1.14).

\begin{figure}
\centering
\pandocbounded{\includegraphics[keepaspectratio,alt={LISA clusters for hate speech rates across Santiago. HH clusters (orange) indicate areas with high hate speech rates surrounded by high rates; LL clusters (purple) indicate low rates surrounded by low rates.}]{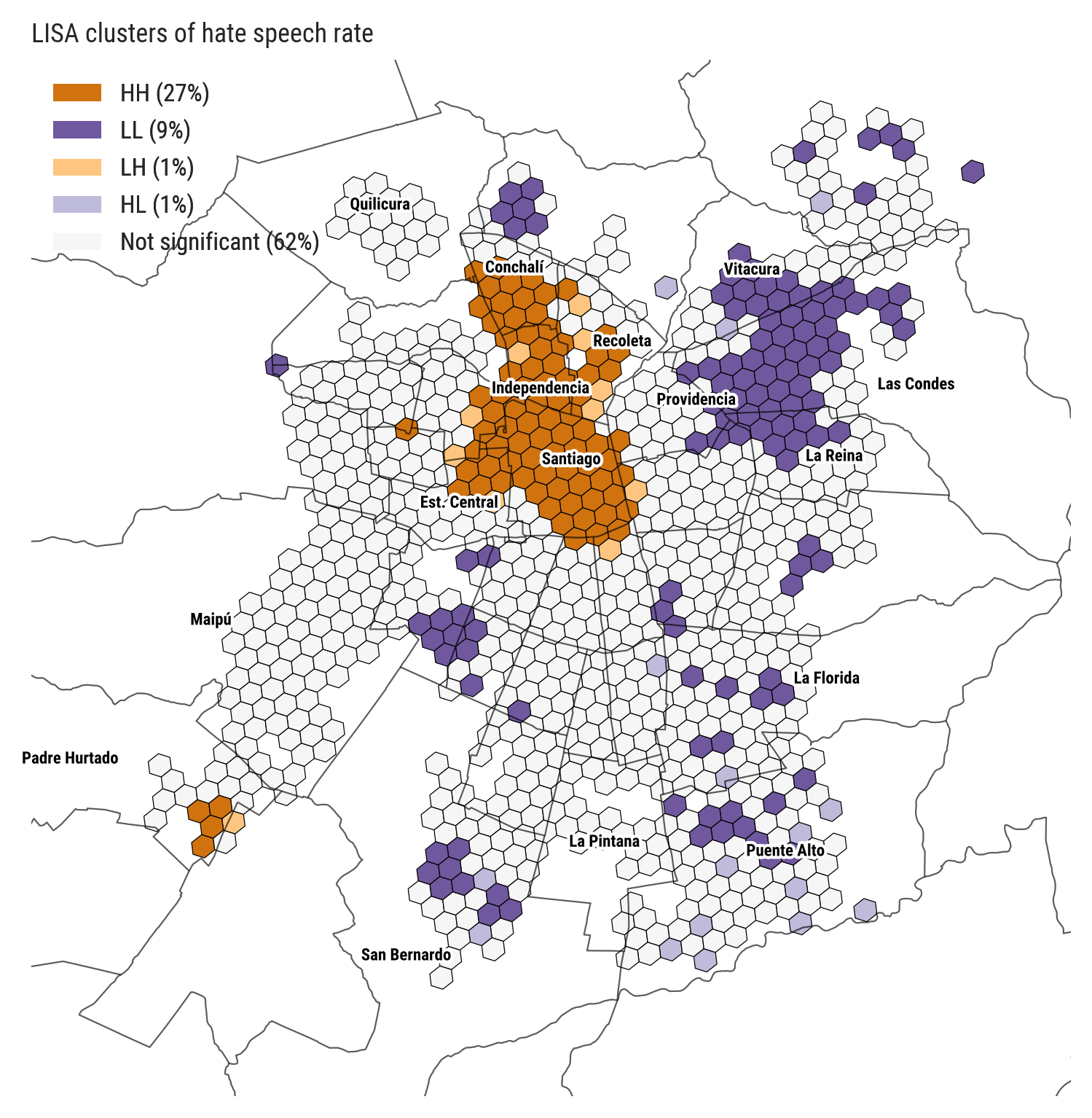}}
\caption{LISA clusters for hate speech rates across Santiago. HH
clusters (orange) indicate areas with high hate speech rates surrounded
by high rates; LL clusters (purple) indicate low rates surrounded by low
rates.}\label{fig:lisa_map}
\end{figure}

Figure \ref{fig:lisa_map} shows the spatial distribution of LISA
clusters. HH clusters (hate speech hotspots) form a contiguous band
across central Santiago, from Conchalí in the north through
Independencia and Recoleta to Santiago and Estación Central. LL clusters
(coldspots) concentrate in the northeastern high-income sector and parts
of the southern periphery. Isolated HL outliers (high rates surrounded
by low) appear within both coldspot zones. Of the 888 local tests, 231
are significant at \(p <\) 0.05; a Benjamini-Hochberg false discovery
rate correction retains 101 of them and preserves the hotspot and
coldspot cores, with the removed cells lying mostly at cluster
boundaries (Appendix E).

Table \ref{tbl:lisa_clusters} summarizes key characteristics by cluster
type. HH clusters contain 27\% of all reports and are in areas with
35.3\% recent migration, more than five times the rate in LL clusters
(6.8\%). HH clusters show elevated hate speech (11.6\% vs.~2.9\% in LL)
and higher migrant mention rates (7.4\% vs.~1.0\%). Established
migration shows no meaningful differentiation across cluster types
(3.29\% in HH vs.~2.89\% in LL). LL clusters concentrate in
high-education areas (48.6\% tertiary education vs.~37.6\% in HH).

\begin{longtable}[]{@{}
  >{\raggedright\arraybackslash}p{(\linewidth - 18\tabcolsep) * \real{0.0921}}
  >{\raggedleft\arraybackslash}p{(\linewidth - 18\tabcolsep) * \real{0.0592}}
  >{\raggedleft\arraybackslash}p{(\linewidth - 18\tabcolsep) * \real{0.0724}}
  >{\raggedleft\arraybackslash}p{(\linewidth - 18\tabcolsep) * \real{0.1184}}
  >{\raggedleft\arraybackslash}p{(\linewidth - 18\tabcolsep) * \real{0.1184}}
  >{\raggedleft\arraybackslash}p{(\linewidth - 18\tabcolsep) * \real{0.1513}}
  >{\raggedleft\arraybackslash}p{(\linewidth - 18\tabcolsep) * \real{0.1053}}
  >{\raggedleft\arraybackslash}p{(\linewidth - 18\tabcolsep) * \real{0.1053}}
  >{\raggedleft\arraybackslash}p{(\linewidth - 18\tabcolsep) * \real{0.0789}}
  >{\raggedleft\arraybackslash}p{(\linewidth - 18\tabcolsep) * \real{0.0987}}@{}}
\caption{\label{tbl:lisa_clusters}LISA cluster
characteristics.}\tabularnewline
\toprule\noalign{}
\begin{minipage}[b]{\linewidth}\raggedright
Cluster type
\end{minipage} & \begin{minipage}[b]{\linewidth}\raggedleft
Reports
\end{minipage} & \begin{minipage}[b]{\linewidth}\raggedleft
Hate rate
\end{minipage} & \begin{minipage}[b]{\linewidth}\raggedleft
Migrant mentions
\end{minipage} & \begin{minipage}[b]{\linewidth}\raggedleft
Recent migration
\end{minipage} & \begin{minipage}[b]{\linewidth}\raggedleft
Established migration
\end{minipage} & \begin{minipage}[b]{\linewidth}\raggedleft
High education
\end{minipage} & \begin{minipage}[b]{\linewidth}\raggedleft
Base education
\end{minipage} & \begin{minipage}[b]{\linewidth}\raggedleft
Avg. likes
\end{minipage} & \begin{minipage}[b]{\linewidth}\raggedleft
Avg. comments
\end{minipage} \\
\midrule\noalign{}
\endfirsthead
\toprule\noalign{}
\begin{minipage}[b]{\linewidth}\raggedright
Cluster type
\end{minipage} & \begin{minipage}[b]{\linewidth}\raggedleft
Reports
\end{minipage} & \begin{minipage}[b]{\linewidth}\raggedleft
Hate rate
\end{minipage} & \begin{minipage}[b]{\linewidth}\raggedleft
Migrant mentions
\end{minipage} & \begin{minipage}[b]{\linewidth}\raggedleft
Recent migration
\end{minipage} & \begin{minipage}[b]{\linewidth}\raggedleft
Established migration
\end{minipage} & \begin{minipage}[b]{\linewidth}\raggedleft
High education
\end{minipage} & \begin{minipage}[b]{\linewidth}\raggedleft
Base education
\end{minipage} & \begin{minipage}[b]{\linewidth}\raggedleft
Avg. likes
\end{minipage} & \begin{minipage}[b]{\linewidth}\raggedleft
Avg. comments
\end{minipage} \\
\midrule\noalign{}
\endhead
\bottomrule\noalign{}
\endlastfoot
HH (hotspots) & 147,314 & 11.6\% & 7.35\% & 35.3\% & 3.29\% & 37.6\% &
35.6\% & 1.87 & 6.64 \\
LL (coldspots) & 53,805 & 2.9\% & 1.03\% & 6.8\% & 2.89\% & 48.6\% &
23.0\% & 1.74 & 7.01 \\
HL (outliers) & 4,750 & 6.7\% & 2.00\% & 6.6\% & 2.29\% & 34.0\% &
34.3\% & 1.36 & 7.62 \\
LH (outliers) & 4,867 & 4.4\% & 3.94\% & 21.5\% & 3.38\% & 33.5\% &
38.8\% & 1.69 & 7.21 \\
Not significant & 326,717 & 4.9\% & 2.08\% & 9.5\% & 1.80\% & 33.3\% &
37.6\% & 1.73 & 8.14 \\
\end{longtable}

Hate speech hotspots appear in areas with mixed educational composition
that also experienced high recent migration, rather than in the most or
least educated areas. Within HH clusters, we split cells at the median
high education level (25.4\%) and identify two profiles. Table
\ref{tbl:hh_split} shows characteristics of HH areas above and below
that threshold.

\begin{longtable}[]{@{}
  >{\raggedright\arraybackslash}p{(\linewidth - 18\tabcolsep) * \real{0.0861}}
  >{\raggedleft\arraybackslash}p{(\linewidth - 18\tabcolsep) * \real{0.0596}}
  >{\raggedleft\arraybackslash}p{(\linewidth - 18\tabcolsep) * \real{0.0728}}
  >{\raggedleft\arraybackslash}p{(\linewidth - 18\tabcolsep) * \real{0.1192}}
  >{\raggedleft\arraybackslash}p{(\linewidth - 18\tabcolsep) * \real{0.1192}}
  >{\raggedleft\arraybackslash}p{(\linewidth - 18\tabcolsep) * \real{0.1523}}
  >{\raggedleft\arraybackslash}p{(\linewidth - 18\tabcolsep) * \real{0.1060}}
  >{\raggedleft\arraybackslash}p{(\linewidth - 18\tabcolsep) * \real{0.1060}}
  >{\raggedleft\arraybackslash}p{(\linewidth - 18\tabcolsep) * \real{0.0795}}
  >{\raggedleft\arraybackslash}p{(\linewidth - 18\tabcolsep) * \real{0.0993}}@{}}
\caption{\label{tbl:hh_split}HH clusters split at median education level
(25.4\%).}\tabularnewline
\toprule\noalign{}
\begin{minipage}[b]{\linewidth}\raggedright
HH subgroup
\end{minipage} & \begin{minipage}[b]{\linewidth}\raggedleft
Reports
\end{minipage} & \begin{minipage}[b]{\linewidth}\raggedleft
Hate rate
\end{minipage} & \begin{minipage}[b]{\linewidth}\raggedleft
Migrant mentions
\end{minipage} & \begin{minipage}[b]{\linewidth}\raggedleft
Recent migration
\end{minipage} & \begin{minipage}[b]{\linewidth}\raggedleft
Established migration
\end{minipage} & \begin{minipage}[b]{\linewidth}\raggedleft
High education
\end{minipage} & \begin{minipage}[b]{\linewidth}\raggedleft
Base education
\end{minipage} & \begin{minipage}[b]{\linewidth}\raggedleft
Avg. likes
\end{minipage} & \begin{minipage}[b]{\linewidth}\raggedleft
Avg. comments
\end{minipage} \\
\midrule\noalign{}
\endfirsthead
\toprule\noalign{}
\begin{minipage}[b]{\linewidth}\raggedright
HH subgroup
\end{minipage} & \begin{minipage}[b]{\linewidth}\raggedleft
Reports
\end{minipage} & \begin{minipage}[b]{\linewidth}\raggedleft
Hate rate
\end{minipage} & \begin{minipage}[b]{\linewidth}\raggedleft
Migrant mentions
\end{minipage} & \begin{minipage}[b]{\linewidth}\raggedleft
Recent migration
\end{minipage} & \begin{minipage}[b]{\linewidth}\raggedleft
Established migration
\end{minipage} & \begin{minipage}[b]{\linewidth}\raggedleft
High education
\end{minipage} & \begin{minipage}[b]{\linewidth}\raggedleft
Base education
\end{minipage} & \begin{minipage}[b]{\linewidth}\raggedleft
Avg. likes
\end{minipage} & \begin{minipage}[b]{\linewidth}\raggedleft
Avg. comments
\end{minipage} \\
\midrule\noalign{}
\endhead
\bottomrule\noalign{}
\endlastfoot
HH (high edu) & 118,302 & 11.7\% & 7.1\% & 37.0\% & 3.27\% & 41.9\% &
32.5\% & 2.00 & 6.19 \\
HH (low edu) & 29,012 & 11.4\% & 8.3\% & 28.4\% & 3.40\% & 20.3\% &
48.5\% & 1.35 & 8.50 \\
\end{longtable}

Hate speech rates are similar in both subgroups (11.7\% and 11.4\%),
even though the low-education subgroup has less recent migration (28.4\%
vs 37.0\%). Migrant mentions are more frequent in lower-education areas
(8.3\% vs 7.1\%). Engagement patterns also differ: high-education areas
receive more likes per report (2.00 vs 1.35) while low-education areas
generate more comments (8.50 vs 6.19).

To assess whether the LISA clusters mediate the regression results, we
re-estimate the hate speech model with cluster fixed effects. Migrant
mention remains stable (\(\beta =\) 2.23 vs.~2.24 in the base model),
while recent migration is attenuated but significant (\(\beta =\) 0.49,
\(p =\) 0.030 vs.~\(\beta =\) 1.54). This attenuation indicates that
part of the recent migration association is absorbed by cluster
membership. The LL cluster shows negative coefficients relative to HH
(\(\beta =\) -0.78, \(p <\) 0.001), which matches the descriptive
pattern.

The classifier over-flags reports that mention migrants (Appendix A),
which could inflate the area-level associations. Re-estimating the model
and the LISA analysis with only the reports that do not mention migrants
leaves the results unchanged: the recent migration coefficient is 1.63
(vs.~1.54), cell-level hate rates correlate at 0.94 with the full-sample
rates, and 86\% of hotspot cells retain their label (Appendix E).

Results are robust to classification threshold variation, differential
classifier error, spatial model specification, resolution choice, rate
smoothing, multiple testing correction, temporal controls, and
institutional platform adoption (Appendices C, D, and E).

\section{Discussion}\label{discussion}

We have shown how anti-immigration sentiment can become embedded in
everyday digital claims about urban disorder. Reports that mention
migrants are more likely to contain hate speech than otherwise similar
reports, and hate speech is spatially concentrated in neighborhoods with
larger shares of recent migrants. Hate speech and migrant mentions also
receive higher platform engagement, although reports that combine both
features are not amplified beyond their separate effects. Together,
these results describe exclusionary discourse in SOSAFE as a repeated,
place-specific, and platform-mediated pattern through which urban
problems can become ethnicized or nationalized.

Citizen reporting platforms make this process observable in ways that
surveys, interviews, and aggregate contextual data rarely can. SOSAFE
reports link text, time, location, incident type, and engagement. This
lets us examine whether anti-immigration sentiment exists, where it
appears, which urban problems it becomes attached to, and how it
circulates. The contribution is therefore both methodological and
substantive. Citizen reporting platforms are active infrastructures
through which residents categorize incidents, make claims visible to
others, and attach urban problems to migrant identity.

These results describe digital bordering as a place-based and
platform-mediated process. In SOSAFE, bordering operates through local
digital complaints rather than through national policy or exclusion from
physical territory. It appears when residents use incident reports to
mark migrant groups as out of place and to link them with disorder or
insecurity in particular neighborhoods. The spatial clustering of hate
speech shows that these framings concentrate where recent migration is
most visible, which ties digital bordering to the local interpretation
of demographic transition.

Our engagement results add a governance dimension to this argument. Hate
speech and migrant mentions are each associated with higher likes and
comments. Exclusionary or identity-marked reports attract more reactions
than otherwise similar content. At the same time, the negative
interaction between hate speech and migrant mentions shows that
migrant-related hate speech is not uniquely amplified beyond the
separate effects of hostility and migrant reference. The risk is
therefore that individual hateful posts attract attention, and that
repeated identity-linked complaints can become more visible as part of
the platform's ordinary engagement dynamics.

Our findings suggest that platform governance should evaluate both
individual reports and aggregate spatial patterns. Moderation systems
that remove or flag isolated abusive posts may miss repeated
place-specific associations between migrant groups and disorder,
especially when individual reports appear as routine neighborhood
complaints. Citizen reporting platforms could combine hate speech
detection with periodic spatial audits and prompts that discourage
unnecessary nationality references. They could also adopt category
designs that focus attention on the reported incident rather than group
identity, and visibility rules that prevent sensitive content from
gaining prominence solely through engagement.

For urban policy, our findings suggest that integration efforts should
prioritize neighborhoods where demographic change is recent. The
distinction between established and recent migration suggests that
interventions should target the process of neighborhood change rather
than the presence of migrants. Community programs that support positive
contact during transition could address intergroup anxiety before it
solidifies into hostility. Within these areas, hate speech occurs at
similar rates in cells with different educational composition, so
exclusionary discourse is not confined to one type of neighborhood.
Explicit nationality references are somewhat more frequent in
lower-education areas, so measures that target only explicit hate speech
may miss subtler forms of exclusion.

Key limitations moderate the scope of our conclusions. First, SOSAFE
users are not representative of Santiago's population. They likely
overrepresent smartphone owners, residents concerned about neighborhood
issues, and groups more likely to use participatory security tools. Our
findings therefore describe the discourse of SOSAFE users rather than
attitudes in the general population. Second, SOSAFE may filter content
before publication. If moderation removes hateful content uniformly, our
estimates are a lower bound of the relationship between migrant mentions
and hate speech; if moderation intensity varies across municipalities,
the direction of the bias is less clear. Third, the classifier is noisy
at the level of individual reports: at deployment prevalence its
precision is 0.15, and its false positives concentrate in reports that
mention migrants (Appendix A). Our conclusions do not rest on individual
labels: the analysis uses aggregate rates, the central association holds
under human labels in a blind random sample (relative risk 9.7 vs.~7.48
with classifier labels), and the area-level results persist when we
exclude all migrant-mention reports (Appendix E). Fourth, keyword
matching identifies explicit migrant mentions but may miss indirect
references or coded language. Fifth, reports are anonymous and carry no
user identifier, so we cannot tell whether a hotspot reflects many
residents or a few prolific reporters; the cluster analysis describes
places, not people.

Future research should examine whether similar patterns occur in other
Chilean and Latin American cities, especially where migrant composition,
platform adoption, and local security politics differ. Qualitative
analysis of report content could clarify how nationality markers become
attached to disorder, insecurity, and belonging in specific local
contexts. More broadly, the findings show that integration policy and
platform governance cannot be treated as separate domains. As
participatory security infrastructures become more common, they will
shape how residents report urban problems and how neighborhood change is
interpreted and made visible.

\section{Conclusions}\label{conclusions}

This study examines how a citizen reporting platform (SOSAFE in
Santiago, Chile) functions as a place where residents produce and
circulate exclusionary discourse that marks migrants as out of place. We
analyze approximately 553,000 geolocated citizen reports and find that
migrant mentions are associated with hate speech, that hate speech
concentrates in areas with recent demographic change rather than in
established migrant communities, and that exclusionary content achieves
greater engagement. These findings are robust to classification
threshold variation, classifier error, spatial specification, temporal
controls, and institutional platform adoption.

The spatial analysis shows a geography of exclusion. Hate speech
hotspots coincide with neighborhoods where recent migrants comprise over
a third of the population, while coldspots appear in high-education
sectors with minimal recent migration. Within hotspots, hate speech
occurs at similar rates in areas with different educational composition,
while explicit nationality references are more frequent in
lower-education areas. These patterns show how participatory security
infrastructures can become channels that politicize migrant presence and
reinforce who belongs where.

Understanding how exclusionary discourse operates requires attention to
platform affordances (engagement metrics, visibility rules, category
systems), which may influence which concerns become visible and how
certain populations come to be framed as problems. For urban
policymakers, the concentration of exclusionary discourse in areas of
recent demographic change suggests that integration efforts should
target the process of neighborhood change. The goal is to facilitate
positive contact during periods of uncertainty, before attitudes
solidify.

As cities worldwide become more diverse, participatory security
infrastructures will play a larger role in mediating how residents
negotiate belonging. Our findings suggest that, without deliberate
design choices, neighborhood reporting practices can sustain repeated
associations between migrant groups and urban disorder.

\section{References}\label{references}

\protect\phantomsection\label{refs}


\begin{CSLReferences}{1}{0}
\bibitem[\citeproctext]{ref-Abadie2017When}
Abadie, A., Athey, S., Imbens, G., \& Wooldridge, J. (2017). When should
you adjust standard errors for clustering? \emph{PSN: Econometrics}.
\url{https://doi.org/10.3386/w24003}

\bibitem[\citeproctext]{ref-allport1954nature}
Allport, G. W., Clark, K., \& Pettigrew, T. (1954). \emph{The nature of
prejudice}.

\bibitem[\citeproctext]{ref-anselin1995local}
Anselin, L. (1995). Local indicators of spatial association--{LISA}.
\emph{Geographical Analysis}, \emph{27}(2), 93--115.
https://doi.org/\url{https://doi.org/10.1111/j.1538-4632.1995.tb00338.x}

\bibitem[\citeproctext]{ref-anselin2014modern}
Anselin, L., \& Rey, S. J. (2014). \emph{Modern spatial econometrics in
practice: A guide to {GeoDa}, {GeoDaSpace} and {PySAL}}. Chicago, IL:
GeoDa Press LLC.

\bibitem[\citeproctext]{ref-arango2024hate}
Arango Monnar, A., Perez Rojas, J., \& Poblete Labra, B. (2024).
Cross-lingual hate speech detection using domain-specific word
embeddings. \emph{PLOS ONE}, \emph{19}(7), 1--18.
\url{https://doi.org/10.1371/journal.pone.0306521}

\bibitem[\citeproctext]{ref-Armstrong2021Challenges}
Armstrong, C., Poorthuis, A., Zook, M., Ruths, D., \& Soehl, T. (2021).
Challenges when identifying migration from geo-located {T}witter data.
\emph{EPJ Data Science}, \emph{10}.
\url{https://doi.org/10.1140/epjds/s13688-020-00254-7}

\bibitem[\citeproctext]{ref-bail2018exposure}
Bail, C. A., Argyle, L. P., Brown, T. W., Bumpus, J. P., Chen, H.,
Hunzaker, M. F., \ldots{} Volfovsky, A. (2018). Exposure to opposing
views on social media can increase political polarization.
\emph{Proceedings of the National Academy of Sciences}, \emph{115}(37),
9216--9221. \url{https://doi.org/10.1073/pnas.1804840115}

\bibitem[\citeproctext]{ref-banks2021polarizedfeeds}
Banks, A., Calvo, E., Karol, D., \& Telhami, S. (2021).
\#{P}olarized{F}eeds: Three experiments on polarization, framing, and
social media. \emph{The International Journal of Press/Politics},
\emph{26}(3), 609--634. \url{https://doi.org/10.1177/1940161220940964}

\bibitem[\citeproctext]{ref-basile2019semeval}
Basile, V., Bosco, C., Fersini, E., Debora, N., Patti, V., Pardo, F. M.
R., et al.others. (2019). Semeval-2019 task 5: Multilingual detection of
hate speech against immigrants and women in {T}witter. \emph{13th
International Workshop on Semantic Evaluation}, 54--63. Association for
Computational Linguistics. \url{https://doi.org/10.18653/v1/S19-2007}

\bibitem[\citeproctext]{ref-bloch2022aversive}
Bloch, S. (2022). Aversive racism and community-instigated policing: The
spatial politics of {N}extdoor. \emph{Environment and Planning C:
Politics and Space}, \emph{40}(1), 260--278.
\url{https://doi.org/10.1177/23996544211019754}

\bibitem[\citeproctext]{ref-bonhomme2022television}
Bonhomme, M., \& Muirhead, A. A. (2022). How television news media
reinforce racialized representations of {H}aitian and {C}olombian
migration in multicultural urban {C}hile. In \emph{Dismantling cultural
borders through social media and digital communications: How networked
communities compromise identity} (pp. 147--184). Springer.
\url{https://doi.org/10.1007/978-3-030-92212-2_6}

\bibitem[\citeproctext]{ref-brown2003teammates}
Brown, K. T., Brown, T. N., Jackson, J. S., Sellers, R. M., \& Manuel,
W. J. (2003). Teammates on and off the field? {C}ontact with black
teammates and the racial attitudes of white student athletes.
\emph{Journal of Applied Social Psychology}, \emph{33}(7), 1379--1403.
\url{https://doi.org/10.1111/j.1559-1816.2003.tb01954.x}

\bibitem[\citeproctext]{ref-bucher2018if}
Bucher, T. (2018). \emph{If... Then: Algorithmic power and politics}.
Oxford University Press.

\bibitem[\citeproctext]{ref-burns2000economic}
Burns, P., \& Gimpel, J. G. (2000). Economic insecurity, prejudicial
stereotypes, and public opinion on immigration policy. \emph{Political
Science Quarterly}, \emph{115}(2), 201--225.

\bibitem[\citeproctext]{ref-calderon2020topic}
Calderón, C. A., Vega, G. de la, \& Herrero, D. B. (2020). Topic
modeling and characterization of hate speech against immigrants on
{T}witter around the emergence of a far-right party in {S}pain.
\emph{Social Sciences}, \emph{9}(11), 188.
\url{https://doi.org/10.3390/socsci9110188}

\bibitem[\citeproctext]{ref-cameron1986econometric}
Cameron, A. C., \& Trivedi, P. K. (1986). Econometric models based on
count data. Comparisons and applications of some estimators and tests.
\emph{Journal of Applied Econometrics}, \emph{1}(1), 29--53.
\url{https://doi.org/10.1002/jae.3950010104}

\bibitem[\citeproctext]{ref-CaneteCFP2020}
Cañete, J., Chaperon, G., Fuentes, R., Ho, J.-H., Kang, H., \& Pérez, J.
(2020). Spanish pre-trained {BERT} model and evaluation data.
\emph{PML4DC at ICLR 2020}.

\bibitem[\citeproctext]{ref-castles2017international}
Castles, S. (2017). International migration at a crossroads. In
\emph{The politics of citizenship in immigrant democracies} (pp.
89--106). Routledge.

\bibitem[\citeproctext]{ref-Cesare2016Promises}
Cesare, N. L., Lee, H., McCormick, T., Spiro, E. S., \& Zagheni, E.
(2016). Promises and pitfalls of using digital traces for demographic
research. \emph{Demography}, \emph{55}, 1979--1999.
\url{https://doi.org/10.1007/s13524-018-0715-2}

\bibitem[\citeproctext]{ref-Chi2020A}
Chi, G., Lin, F., Chi, G., \& Blumenstock, J. (2020). A general approach
to detecting migration events in digital trace data. \emph{PLoS ONE},
\emph{15}. \url{https://doi.org/10.1371/journal.pone.0239408}

\bibitem[\citeproctext]{ref-choksi2024under}
Choksi, M. Z., Aubin Le Quéré, M., Lloyd, T., Tao, R., Grimmelmann, J.,
\& Naaman, M. (2024). Under the (neighbor)hood: Hyperlocal surveillance
on {N}extdoor. \emph{Proceedings of the 2024 CHI Conference on Human
Factors in Computing Systems}, 1--22.
\url{https://doi.org/10.1145/3613904.3641967}

\bibitem[\citeproctext]{ref-comandini2019impossible}
Comandini, G., \& Patti, V. (2019). An impossible dialogue! Nominal
utterances and populist rhetoric in an {I}talian {T}witter corpus of
hate speech against immigrants. \emph{Third Workshop on Abusive Language
Online}, 163--171. Association for Computational Linguistics.

\bibitem[\citeproctext]{ref-Crespo2020On}
Crespo, R., \& Hernandez, I. (2020). On the spatially explicit {G}ini
coefficient: The case study of {C}hile--a high-income developing
country. \emph{Letters in Spatial and Resource Sciences}, \emph{13},
37--47. \url{https://doi.org/10.1007/s12076-020-00243-4}

\bibitem[\citeproctext]{ref-dammert2019crimen}
Dammert, L., \& Sandoval, R. (2019). Crimen, inseguridad y migraci{ó}n:
De la percepci{ó}n a la realidad. \emph{Migraci{ó}n En Chile: Evidencia
y Mitos de Una Nueva Realidad}, 199--230.

\bibitem[\citeproctext]{ref-dennison2018public}
Dennison, J., \& Dražanová, L. (2018). \emph{Public attitudes on
migration: Rethinking how people perceive migration: An analysis of
existing opinion polls in the {E}uro-{M}editerranean region}. European
University Institute. Retrieved from European University Institute
website: \url{https://hdl.handle.net/1814/62348}

\bibitem[\citeproctext]{ref-dennison2019rising}
Dennison, J., \& Geddes, A. (2019). A rising tide? The salience of
immigration and the rise of anti-immigration political parties in
{W}estern {E}urope. \emph{The Political Quarterly}, \emph{90}(1),
107--116. \url{https://doi.org/10.1111/1467-923X.12620}

\bibitem[\citeproctext]{ref-dromundo2025sinodio}
Dromundo, A. (2025a). \emph{Sin{O}dio {BETO}: Detector de discurso de
odio para {E}spañol {L}atinoamericano}. HuggingFace. Retrieved from
\url{https://huggingface.co/antonn-dromundo/SinOdio-BETO-HateSpeech-Detector-v3}

\bibitem[\citeproctext]{ref-dromundo2024sinodio_latam}
Dromundo, A. (2025b). \emph{{SinOdio-LATAM-Regional-HateSpeech}: Hate
speech detection dataset for {L}atin {A}merican {S}panish} {[}Data
set{]}.
\url{https://huggingface.co/datasets/antonn-dromundo/SinOdio-LATAM-Regional-HateSpeech};
Hugging Face.

\bibitem[\citeproctext]{ref-esses2002immigration}
Esses, V. M., Dovidio, J. F., Jackson, L. M., \& Armstrong, T. L.
(2002). The immigration dilemma: The role of perceived group
competition, ethnic prejudice, and national identity. \emph{Journal of
Social Issues}, \emph{57}(3), 389--412.
\url{https://doi.org/10.1111/0022-4537.00220}

\bibitem[\citeproctext]{ref-fortuna2018survey}
Fortuna, P., \& Nunes, S. (2018). A survey on automatic detection of
hate speech in text. \emph{ACM Computing Surveys (CSUR)}, \emph{51}(4),
1--30. \url{https://doi.org/10.1145/3232676}

\bibitem[\citeproctext]{ref-freier2019regional}
Freier, L. F., \& Parent, N. (2019). The regional response to the
{V}enezuelan exodus. \emph{Current History}, \emph{118}(805), 56--61.

\bibitem[\citeproctext]{ref-freire2021framework}
Freire-Vidal, Y., Graells-Garrido, E., \& Rowe, F. (2021). A framework
to understand attitudes towards immigration through {T}witter.
\emph{Applied Sciences}, \emph{11}(20), 9689.
\url{https://doi.org/10.3390/app11209689}

\bibitem[\citeproctext]{ref-grow2021reliable}
Grow, A., Perrotta, D., Del Fava, E., Cimentada, J., Rampazzo, F.,
Gil-Clavel, S., \ldots{} Weber, I. (2021). \emph{How reliable is
{F}acebook's advertising data for use in social science research?
Insights from a cross-national online survey}. Max Planck Institute for
Demographic Research, Rostock, Germany.
https://doi.org/\url{https://dx.doi.org/10.4054/MPIDR-WP-2021-006}

\bibitem[\citeproctext]{ref-hanson2007public}
Hanson, G. H., Scheve, K., \& Slaughter, M. J. (2007). Public finance
and individual preferences over globalization strategies.
\emph{Economics \& Politics}, \emph{19}(1), 1--33.
\url{https://doi.org/10.1111/j.1468-0343.2007.00300.x}

\bibitem[\citeproctext]{ref-hopkins2010politicized}
Hopkins, D. J. (2010). Politicized places: Explaining where and when
immigrants provoke local opposition. \emph{American Political Science
Review}, \emph{104}(1), 40--60.
\url{https://doi.org/10.1017/S0003055409990360}

\bibitem[\citeproctext]{ref-censo-2024}
Instituto Nacional de Estadísticas. (2025). \emph{Censo de población y
vivienda 2024}. \url{https://www.ine.gob.cl/censo2024}.

\bibitem[\citeproctext]{ref-iqbal2023lady}
Iqbal, W., Ghafouri, V., Tyson, G., Suarez-Tangil, G., \& Castro, I.
(2023). Lady and the tramp {N}extdoor: Online manifestations of
real-world inequalities in the {N}extdoor social network.
\emph{Proceedings of the International AAAI Conference on Web and Social
Media}, \emph{17}, 399--410.
\url{https://doi.org/10.1609/icwsm.v17i1.22155}

\bibitem[\citeproctext]{ref-jolly2014xenophobia}
Jolly, S. K., \& DiGiusto, G. M. (2014). Xenophobia and immigrant
contact: {F}rench public attitudes toward immigration. \emph{The Social
Science Journal}, \emph{51}(3), 464--473.
\url{https://doi.org/10.1016/j.soscij.2013.09.018}

\bibitem[\citeproctext]{ref-keipi2016online}
Keipi, T., Näsi, M., Oksanen, A., \& Räsänen, P. (2016). \emph{Online
hate and harmful content: Cross-national perspectives}. Taylor \&
Francis. \url{https://doi.org/10.4324/9781315628370}

\bibitem[\citeproctext]{ref-kirkpatrick1983optimization}
Kirkpatrick, S., Gelatt Jr, C. D., \& Vecchi, M. P. (1983). Optimization
by simulated annealing. \emph{Science}, \emph{220}(4598), 671--680.
\url{https://doi.org/10.1126/science.220.4598.671}

\bibitem[\citeproctext]{ref-kopstein2009does}
Kopstein, J. S., \& Wittenberg, J. (2009). Does familiarity breed
contempt? {I}nter-ethnic contact and support for illiberal parties.
\emph{The Journal of Politics}, \emph{71}(2), 414--428.
\url{https://doi.org/10.1017/S0022381609090367}

\bibitem[\citeproctext]{ref-kopytowska2015discourse}
Kopytowska, M. (2015). Discourse of hate and radicalism in action.
\emph{Journal of Language Aggression \& Conflict}, \emph{3}(1).
\url{https://doi.org/10.1075/jlac.3.1.001ed}

\bibitem[\citeproctext]{ref-lambright2019digital}
Lambright, K. (2019). Digital redlining: The {N}extdoor app and the
neighborhood of make-believe. \emph{Cultural Critique}, \emph{103},
84--90. \url{https://doi.org/10.5749/culturalcritique.103.2019.0084}

\bibitem[\citeproctext]{ref-larsson2017first}
Larsson, S. (2017). A first line of defence? Vigilant surveillance,
participatory policing and the reporting of {``suspicious''} activity.
\emph{Surveillance \& Society}, \emph{15}(1), 94--107.
https://doi.org/\url{https://doi.org/10.24908/ss.v15i1.5342}

\bibitem[\citeproctext]{ref-lawrence2015crossing}
Lawrence, D. (2015). Crossing the {C}ordillera: Immigrant attributes and
{C}hilean attitudes. \emph{Latin American Research Review},
\emph{50}(4), 154--177. \url{https://doi.org/10.1353/lar.2015.0058}

\bibitem[\citeproctext]{ref-openshaw1984modifiable}
Openshaw, S. (1984). The {M}odifiable {A}real {U}nit {P}roblem.
\emph{Concepts and Techniques in Modern Geography}.

\bibitem[\citeproctext]{ref-pariser2011filter}
Pariser, E. (2011). \emph{The {F}ilter {B}ubble: How the new
personalized web is changing what we read and how we think}. Penguin.

\bibitem[\citeproctext]{ref-parker2024new}
Parker, M., \& Dodge, M. (2024). The new neighborhood watch: An
exploratory study of the {N}extdoor app and crime narratives.
\emph{International Journal of Criminology and Sociology}, \emph{13},
43--54. \url{https://doi.org/10.6000/1929-4409.2024.13.04}

\bibitem[\citeproctext]{ref-perez2022robertuito}
Pérez, J. M., Furman, D. A., Alonso Alemany, L., \& Luque, F. M. (2022).
{RoBERTuito}: A pre-trained language model for social media text in
{S}panish. \emph{Proceedings of the Language Resources and Evaluation
Conference}, 7235--7243. Marseille, France: European Language Resources
Association. Retrieved from
\url{https://aclanthology.org/2022.lrec-1.785}

\bibitem[\citeproctext]{ref-perez2021twitter}
Pérez-Arredondo, C., \& Graells-Garrido, E. (2021). Twitter and
abortion: Online hate against pro-choice female politicians in {C}hile.
\emph{Journal of Language Aggression and Conflict}, \emph{9}(1),
127--154. \url{https://doi.org/10.1075/jlac.00056.per}

\bibitem[\citeproctext]{ref-perez2025immigrant}
Pérez-Arredondo, C., Ivanova, A., \& Graells-Garrido, E. (2025).
Immigrant-friendly stakeholders and the use of {T}witter as a digital
third space in {C}hile: Is that so? \emph{Journal of Language and
Discrimination}, \emph{9}(1), 57--83.
\url{https://doi.org/10.3138/jld-2025-0503}

\bibitem[\citeproctext]{ref-pettigrew2006meta}
Pettigrew, T. F., \& Tropp, L. R. (2006). A meta-analytic test of
intergroup contact theory. \emph{Journal of Personality and Social
Psychology}, \emph{90}(5), 751.

\bibitem[\citeproctext]{ref-putnam2007pluribus}
Putnam, R. D. (2007). E pluribus unum: Diversity and community in the
twenty-first century the 2006 {J}ohan {S}kytte {P}rize {L}ecture.
\emph{Scandinavian Political Studies}, \emph{30}(2), 137--174.

\bibitem[\citeproctext]{ref-rowe2021using}
Rowe, F., Mahony, M., Graells-Garrido, E., Rango, M., \& Sievers, N.
(2021). Using {T}witter to track immigration sentiment during early
stages of the {COVID}-19 pandemic. \emph{Data \& Policy}, \emph{3}, e36.
\url{https://doi.org/10.1017/dap.2021.38}

\bibitem[\citeproctext]{ref-sanguinetti2018italian}
Sanguinetti, M., Poletto, F., Bosco, C., Patti, V., \& Stranisci, M.
(2018). An {I}talian {T}witter corpus of hate speech against immigrants.
\emph{Proceedings of the Eleventh International Conference on Language
Resources and Evaluation (LREC 2018)}.

\bibitem[\citeproctext]{ref-scherman2022influence}
Scherman, A., Etchegaray, N., Pavez, I., \& Grassau, D. (2022). The
influence of media coverage on the negative perception of migrants in
{C}hile. \emph{International Journal of Environmental Research and
Public Health}, \emph{19}(13), 8219.
\url{https://doi.org/10.3390/ijerph19138219}

\bibitem[\citeproctext]{ref-scheve2001labor}
Scheve, K. F., \& Slaughter, M. J. (2001). Labor market competition and
individual preferences over immigration policy. \emph{Review of
Economics and Statistics}, \emph{83}(1), 133--145.

\bibitem[\citeproctext]{ref-seabold2010statsmodels}
Seabold, S., Perktold, J., et al. (2010). Statsmodels: Econometric and
statistical modeling with {P}ython. \emph{SciPy}, \emph{7}(1), 92--96.

\bibitem[\citeproctext]{ref-stephan1985intergroup}
Stephan, W. G., \& Stephan, C. W. (1985). Intergroup anxiety.
\emph{Journal of Social Issues}, \emph{41}(3), 157--175.
https://doi.org/\url{https://doi.org/10.1111/j.1540-4560.1985.tb01134.x}

\bibitem[\citeproctext]{ref-tironi2021circulation}
Tironi, M., \& Albornoz, C. (2021). The circulation of the {S}mart
{C}ity imaginary in the {C}hilean context: A case study of a
collaborative platform for governing security. In \emph{Smart cities for
technological and social innovation} (pp. 195--215). Elsevier.
\url{https://doi.org/10.1016/B978-0-12-818886-6.00011-3}

\bibitem[\citeproctext]{ref-h3geo}
Uber Technologies. (2026). \emph{{H}3}. \url{https://h3geo.org/}.

\bibitem[\citeproctext]{ref-UNHateSpeech}
United Nations. (n.d.). \emph{What is hate speech?}
\url{https://www.un.org/en/hate-speech/understanding-hate-speech/what-is-hate-speech}.

\bibitem[\citeproctext]{ref-vosoughi2018spread}
Vosoughi, S., Roy, D., \& Aral, S. (2018). The spread of true and false
news online. \emph{Science}, \emph{359}(6380), 1146--1151.
\url{https://doi.org/10.1126/science.aap9559}

\bibitem[\citeproctext]{ref-yuval2019bordering}
Yuval-Davis, N., Wemyss, G., \& Cassidy, K. (2019). \emph{Bordering}.
Polity Press.

\end{CSLReferences}
\end{document}